\documentclass[sigconf,natbib,nonacm]{acmart} %

\input{sigir25-rag-crowdsourcing-frame.sty}
\graphicspath{{./figures}}

\newcommand{\bswebis}[1][ ]{} %
\newcommand{\bswebisuc}[1][ ]{} %

\begin{document}
\title{The Viability of Crowdsourcing for RAG Evaluation}

\settopmatter{authorsperrow=4}

\author[L.~Gienapp]{Lukas Gienapp}
\affiliation{
  \institution{\mbox{\kern-1em Leipzig University and~ScaDS.AI\hspace{-11pt}}}
  \city{Leipzig} 
  \country{Germany}
}
\orcid{0000-0001-5707-3751}

\author[T.~Hagen]{Tim Hagen}
\affiliation{
  \institution{University of Kassel and~hessian.AI}
  \city{Kassel}
  \country{Germany}
}
\orcid{0009-0000-4854-7249}

\author[M.~Fr{\"o}be]{Maik Fr{\"o}be}
\affiliation{
  \institution{Friedrich-Schiller-Universit{\"a}t Jena}
  \city{Jena}
  \country{Germany}
}
\orcid{0000-0002-1003-981X}

\author[M.~Hagen]{Matthias Hagen}
\affiliation{
  \institution{Friedrich-Schiller-Universit{\"a}t Jena}
  \city{Jena}
  \country{Germany}
}
\orcid{0000-0002-9733-2890}

\author[B.~Stein]{Benno Stein}
\affiliation{
  \institution{\mbox{\kern-0.5em Bauhaus-Universit{\"a}t~Weimar\hspace{-2pt}}}
  \city{Weimar}
  \country{Germany}
}
\orcid{0000-0001-9033-2217}

\author[M.~Potthast]{Martin Potthast}
\affiliation{
  \institution{University of Kassel, hessian.AI, and ScaDS.AI}
  \city{Kassel}
  \country{Germany}
}
\orcid{0000-0003-2451-0665}

\author[H.~Scells]{Harrisen Scells}
\affiliation{
  \institution{University of Kassel and~hessian.AI}
  \city{Kassel}
  \country{Germany}
}
\orcid{0000-0001-9578-7157}

\renewcommand{\shortauthors}{Lukas Gienapp et al.}

\begin{abstract}
How good are humans at writing and judging responses in retrieval-augmented generation~(RAG) scenarios? To answer this question, we investigate the efficacy of crowdsourcing for RAG through two complementary studies: response writing and response utility judgment. We present the \bswebis Crowd RAG Corpus~2025 (\bswebis[-]CrowdRAG-25), which consists of 903~human-written and 903~LLM-generated responses for the 301~topics of the TREC RAG'24~track, across the three discourse styles `bulleted list', `essay', and `news'. For a selection of 65~topics, the corpus further contains 47,320~pairwise human judgments and 10,556~pairwise LLM judgments across seven utility dimensions (e.g., coverage and coherence). Our analyses give insights into human writing behavior for RAG and the viability of crowdsourcing for RAG evaluation. Human pairwise judgments provide reliable and cost-effective results compared to LLM-based pairwise or human/LLM-based pointwise judgments, as well as automated comparisons with human-written reference responses. All our data and tools are freely available.%
\footnote{%
  \hbox to 2.5em{Data: }\url{https://zenodo.org/records/14748980} \\
  \phantom{\textsuperscript{1}}\hbox to 2.5em{Code: }\url{https://github.com/webis-de/sigir25-rag-crowdsourcing} \\[10em]
}

\end{abstract}



\copyrightyear{2025}
\acmYear{2025}
\setcopyright{cc}
\setcctype{by}
\acmConference[SIGIR '25]{Proceedings of the 48th International ACM SIGIR Conference on Research and Development in Information Retrieval}{July 13--18, 2025}{Padua, Italy}
\acmBooktitle{Proceedings of the 48th International ACM SIGIR Conference on Research and Development in Information Retrieval (SIGIR '25), July 13--18, 2025, Padua, Italy}\acmDOI{10.1145/3726302.3730093}
\acmISBN{979-8-4007-1592-1/2025/07}

\settopmatter{printacmref=true}

\makeatletter
\gdef\@copyrightpermission{
 \begin{minipage}{0.3\columnwidth}
  \href{https://creativecommons.org/licenses/by/4.0/}{\includegraphics[width=0.90\textwidth]{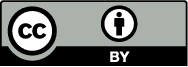}}
 \end{minipage}\hfill
 \begin{minipage}{0.7\columnwidth}
  \href{https://creativecommons.org/licenses/by/4.0/}{This work is licensed under a Creative Commons Attribution International 4.0 License.}
 \end{minipage}
 \vspace{5pt}
}
\makeatother

\maketitle

\section{Introduction}

The introduction of large language models~(LLMs) caused a paradigm shift in information retrieval~(IR). LLMs enabled the development of a new generation of search engines that implement retrieval-augmented generation~(RAG)~\cite{lewis:2020}: Instead of the traditional search engine results page in the form of a ranked list of retrieved documents (list~SERP), RAG systems return a text response in the style of a direct answer~\cite{potthast:2020} (text~SERP~\cite{gienapp:2024}). The goal of RAG systems is to relieve users from browsing search results by synthesizing direct, coherent, and satisfactory answers to their queries, based on information from documents retrieved by a backend search engine.

\begin{table}
\small
\centering
\caption{Key figures of \bswebis[-]CRAGC-25 and the two crowdsourcing studies carried out to compile the corpus.}
\label{table-webis-crowd-rag-corpus-25-key-figures}
\begin{tabular}[t]{@{}l@{\ \ }r@{\hspace{1em}}l@{\ \ }r@{}}
\toprule
\multicolumn{4}{@{}c@{}}{\bf The \bswebis Crowd RAG Corpus 2025} \\
\midrule
\multicolumn{2}{@{}c@{\hspace{1em}}}{RAG Response Writing} & \multicolumn{2}{@{}c@{}}{RAG Response Judgment} \\
\cmidrule(r@{1em}){1-2}\cmidrule{3-4}
  Topics {\scriptsize (TREC RAG'24 \cite{pradeep:2024})} &                                          301 & Topics {\scriptsize (TREC RAG'24 \cite{pradeep:2024})} &                 65 \\
  RAG responses                                          &                                        1,806 & Quality criteria \cite{gienapp:2024}                   &                  7 \\
  \rotatebox[origin=c]{180}{$\Lsh$}~Bullet list style    & \kern-3em 301 \emojihuman\ + 301 \emojimodel & Response pairs                                         &              1,352 \\
  \rotatebox[origin=c]{180}{$\Lsh$}~Essay style          & \kern-3em 301 \emojihuman\ + 301 \emojimodel & \emojihuman\ pairwise judgments                        &             47,320 \\
  \rotatebox[origin=c]{180}{$\Lsh$}~News style           & \kern-3em 301 \emojihuman\ + 301 \emojimodel & \emojimodel\ pairwise judgments                        &             10,556 \\
\cmidrule(r@{1em}){1-2}\cmidrule{3-4}
  Human workers                                          &                                           34 & Crowd judges                                           &                420 \\
  Words per response                                     &                                $\approx$ 270 & Judgments per judge                                    &                112 \\
  Time per response                                      &                            $\approx$ 11 min. & Time per judgment                                      & $\approx$ 0.6 min. \\
  Hourly rate                                            &                            $\approx$ \$14.40 & Hourly rate                                            &  $\approx$ \$13.20 \\
\cmidrule(r@{1em}){1-2}\cmidrule{3-4}
  Cost per response                                      &                                       \$2.89 & Cost per (gold) judgment                               &    (\$0.30) \$0.06 \\
\bottomrule
\end{tabular}
\end{table}

Unlike for traditional search engines, there is currently no com\-munity-wide standard for the evaluation of RAG systems in laboratory settings. Therefore, many new evaluation approaches have been proposed in the last two years (see Section~\ref{related-work}). They can be grouped into two basic paradigms~\cite{gienapp:2024}: \emph{judgment-based} evaluation, inspired by traditional IR evaluation, which ranks systems by assigning explicit utility judgments to their responses, and \emph{reference-based} evaluation, inspired by summarization research, which ranks systems by comparing their generated response to a ground-truth reference using a similarity measure. Despite their past successful use in natural language generation tasks, the former is not very scalable, whereas the latter, apart from a high up-front overhead, still lacks similarity measures that correlate with user preferences~\cite{goyal:2022,zhang:2024}. In both cases, LLMs have been proposed and used to replace human judges~\cite{friel:2024,shahul:2024,upadhyay:2024,pradeep:2024}, to generate reference texts~\cite{kamalloo:2023}, and as new, advanced similarity measures~\cite{li:2024}. The use of LLMs to judge LLM-generated RAG responses, however, has been criticized~\cite{soboroff:2024,clarke:2024}, arguing for humans as the only valid source of utility evidence, possibly supported by LLMs~\cite{faggioli:2023b,dietz:2024a}.

To our knowledge, no systematic investigation of the capabilities and limitations of human-sourced ground-truth data for retrieval-augmented generation has not done so far. We therefore investigate the viability of crowdsourcing as a scalable alternative for RAG evaluation. With cost-effectiveness in mind, we ask how well humans can write reference RAG responses, and how reliable and valid crowdsourced RAG response judgments are, using LLM-based responses, and judgments, respectively, for comparison. As a result, we compile the \bswebis Crowd RAG Corpus~2025 (\bswebis[-]CrowdRAG-25) summarized in Table~\ref{table-webis-crowd-rag-corpus-25-key-figures}. First, for reference-based evaluation, we collect 903~human-written and 903~LLM-generated RAG~responses for all 301~topics of the TREC RAG'24 track~\cite{pradeep:2024}, encompassing 301~responses for each of three potential RAG~discourse styles: bulleted lists, essays, and news. Second, all~1,806 responses are judged by a different group of human crowd workers, and an LLM. We collect a total of 47,320~pairwise human judgments and 10,556~pairwise LLM judgments across seven RAG-specific utility dimensions~\cite{gienapp:2024}. Our analysis compares the writing of humans and LLMs for RAG, and the efficacy of both evaluation paradigms in obtaining reliable and valid results.

\section{Related Work}
\label{related-work}

We review evaluation approaches for retrieval-augmented generation systems and then discuss related work on using crowdsourcing and LLMs as sources of ground-truth in IR and beyond. 

\paragraph{Retrieval-Augmented Generation}
Retrieval-augmented generation~(RAG) has been originally proposed as a means to expand the knowledge recall of LLMs beyond what they internalized during training by conditioning them on documents relevant to a given prompt that are retrieved during generation~\cite{lewis:2020}. However, this implicit form of augmentation at the attention level was quickly complemented by an explicit augmentation at the instruction level~\cite{gienapp:2024}: Rather than just producing statements that correctly reproduce retrieved sources with a high probability, RAG systems are now expected to explicitly cite them. Today's RAG systems replace the traditional listing of sources on a search results page (list SERP) with a summary referencing the retrieved sources (text SERP). \citet{gao:2023} provide an overview of the state of the art in RAG approaches. The shape of RAG responses has quickly arrived at what amounts to a new industry standard. However, no consensus has yet been reached on how RAG systems should be evaluated. In the literature on RAG evaluation, two paradigms can be distinguished: reference-based evaluation and judgment-based evaluation. A key point of discussion is to what extent the evaluation can be automated.

\paragraph{Reference-based Evaluation}
Since RAG responses are a kind of multi-document summary, following summarization evaluation, reference-based evaluation measures the utility of a RAG response by comparing it with a ground-truth response on the same topic~\cite{gienapp:2024}. Corresponding RAG benchmarks have been developed~\cite{lyu:2024,liu:2023,friel:2024,wang:2024,tang:2024,chen:2024}, where the comparison is made using a similarity measure such as BLEU~\cite{post:2018}, ROUGE~\cite{lin:2004}, BERTScore~\cite{zhang:2020}, exact matching, or fine-tuned language models~\cite{li:2024b}. However, reference-based evaluation has been shown to not accurately differentiate the actual effectiveness of systems: For example, for news summarization, \citet{zhang:2024} and \citet{goyal:2022} show that human preferences often do not match reference-based results. For RAG, these measures have not been validated against human preferences.

\paragraph{Judgment-based Evaluation}
Following traditional IR evaluation, judgment-based evaluation assigns explicit utility judgments to RAG responses. \citet{gienapp:2024} review existing work and compile an overview of the different utility dimensions for RAG, grouped under the five top-level dimensions of coherence, coverage, consistency, correctness, and clarity. \citet{hosking:2024} collected human judgments for a generic language generation task and found that undifferentiated judgments tend to be biased against certain utility dimensions. Other studies relying on human judgments focus only on subsets of the utility dimensions~\cite{zhang:2024b,kamalloo:2023}.

\paragraph{LLM-based Evaluation Automation}
As of recently, LLMs are being used to collect query relevance judgments (qrels)~\cite{faggioli:2023b,rahmani:2024,rahmani:2024b,thomas:2024,upadhyay:2024,upadhyay:2024b,macavaney:2023}. Moreover, LLMs have been used for `query utility judgments' (qutils) for RAG responses as well~\cite{friel:2024,shahul:2024,li:2024}. Reference-based ground-truth data is also increasingly produced using LLMs across disciplines, including generating documents~\cite{turkmen:2025} and ground-truth responses~\cite{kamalloo:2023}. LLMs are also used to generate synthetic training data to fine-tune evaluation models~\cite{saad-falcon:2024}. For text generation, LLMs have produced output that human test subjects found hard to distinguish from human-written texts~\cite{clark:2021}, surpassing the latters' quality on specific tasks~\cite{zhang:2024}. However, the use of LLMs to evaluate LLM output has been criticized as circular~\cite{soboroff:2024,clarke:2024}.

\paragraph{Crowdsourcing for Evaluation}
Crowdsourcing has been successfully used in~IR~\cite{alonso:2011} as a source of query relevance judgments~\cite{alonso:2012}. It has been shown to produce reliable judgments~\cite{roitero:2021} in a scalable manner~\cite{alonso:2012} that can be validated with expert judgments~\cite{alonso:2009}. Quality dimensions beyond relevance have also been successfully measured in this way~\cite{gienapp:2020}. However, given that \citeauthor{hosking:2024}'s~\cite{hosking:2024} results suggest that asking a human for a single, overall judgment of a RAG response tends to introduce bias, e.g., against factuality and consistency in responses, differentiating utility may be necessary. Moreover, \citeauthor{hosking:2024} corroborate the findings of multiple studies~\cite{novikova:2018,gienapp:2020,goyal:2022,zhang:2024} that show that pairwise judgments produce reliable outcomes w.r.t. fine-grained text quality dimensions, contrary to pointwise judgments. Nugget-based evaluation~\cite{voorhees:2003,lin:2006} was proposed to simplify pointwise workflows, also for RAG~\cite{pradeep:2024b}, but is primarily intended for relevance, not quality assessment.

Crowdsourcing has also been employed to collect human-written text. For example, \citet{zhang:2024} conducted a crowdsourcing study to evaluate single-document news summarization, where crowd workers formulated reference summaries. \citet{verroios:2014} collected human-written summaries in a multi-stage writing process. \citet{hagen:2016} analyzed writing progress and source usage for open-ended writing tasks with multiple source documents. The viability of crowdsourcing for collecting judgments and text has thus been extensively demonstrated, both in IR and beyond. Crowdsourcing will therefore also be useful for RAG evaluation.

\section{The \texorpdfstring{\bswebis}{} Crowd RAG Corpus 2025}
\label{webis-crowd-rag-corpus-2025}

Following established best practices for crowdsourcing~\cite{alonso:2015,quinn:2017}, we design two crowdsourcing studies to gather RAG responses, query utility judgments, and sufficient evidence to verify the study design.

\subsection{RAG Response Writing}\label{sec:data-collection}

With the first crowdsourcing study, we gather human-written RAG responses for a set of topics, written in three different discourse styles. Similarly, LLM responses are compiled for comparison.

\subsubsection{Topics, Retrieval Results, and Preliminary Steps}

To crowdsource RAG responses, a set of topics is required, where for each topic a ranked list of relevant retrieval results is available. To maximize synergies with existing and future research, we reuse the 301~topics of the recent TREC RAG'24 track~\cite{pradeep:2024}. The track also supplies a document collection, where each document has been pre-processed to extract a total of 113M~passages. One of the tasks at TREC RAG'24 was to retrieve relevant passages for RAG response generation, and we reused the system~\texttt{webis-01} for our study, one of the top most effective at TREC RAG'24 with a focus on recall.%

\subsubsection{Study Design}

For every topic and its top-20 most relevant passages, a crowd worker was tasked with composing a 250~word RAG response. The web-based writing interface we implemented shows a text editor and next to it the 20~retrieved passages as a ranked list. Since this list exceeds the screen height, scrolling through it does not move the editor out of view. A basic JavaScript-based search allows a worker to filter the passages. Workers received writing instructions, an explanation of our study, and guidelines to cite claims taken from passages using a prescribed citation format, with multiple references where applicable. Workers were also prohibited from using language models to complete the task. In this respect, we informed them that we tracked their interactions with our interface including saving the current text version at 300ms~intervals until completion, as well as clicks, copy/paste events, dwell times, key presses, etc., via the BigBro library~\citep{scells:2021}.

As part of our research, we were interested in studying alternative discourse styles of RAG responses. For each topic, we asked three different workers to compose their RAG response in one of the following discourse styles:
\begin{enumerate*}[label=(\arabic*)]
\item
\emph{bullet} list style, listing all the points relevant to the topic as a bulleted list;
\item
\emph{essay} style, starting with a clear thesis, then providing arguments, and finishing with a conclusion; 
 and 
\item
\emph{news} style, starting with the lead, then providing the important details, and lastly adding background information (i.e., the ``inverted pyramid'' scheme of news article).
\end{enumerate*}

\subsubsection{Worker Recruitment}

We recruit workers on the Upwork platform.%
\footnote{\url{www.upwork.com}} 
A detailed job ad was posted to which interested workers could apply. They were hired after manual review of their worker profile, their previous work, and successfully writing a paid response for an example topic. A total of 34~workers were recruited. On average, each worker composed 26~responses.

\subsubsection{Review Process}

We identify four factors that could impact the quality of the collected text responses:
\begin{enumerate*}[label=(\arabic*)]
\item
bad input data, i.e., the topic or the retrieved passages not being suitable to formulate high-quality responses,
\item
workers misunderstanding assignments,
\item
usage of generative models, and
\item
spam.
\end{enumerate*}
We account for these factors through a combination of entry and exit questionnaires, manual checks, and user interaction analysis.

To check for bad input data and task understanding, workers were asked to fill out entry and exit questionnaires. The entry questionnaire asks for a self-assessment on topic knowledge and expertise, and whether a worker considers the topic answerable as well as whether it is controversial. The exit questionnaire asks for how satisfied a worker was with their response, if the provided passages were adequate sources of knowledge on the topic, and whether own prior knowledge was introduced. All questions were answered on either a 1-5~Likert scale or a ternary `yes'/`maybe'/`no' scale. Self-assessed worker knowledge and expertise were highly correlated~($\rho = 0.88$), with expertise (median of~1/5) slightly lower than knowledge (median of~2/5). The quality of passages showed no significant correlation with the self-assessed use of own prior knowledge ($\rho = -0.04$). Prior knowledge use was generally low (assessed~1/5 for~66\% of responses), while result passage quality was consistently judged high (median judgment of~4/5). In~60\% of the responses, workers reported that some of the given passages were omitted. Omissions were not contingent on passage quality, as the proportion of omissions is approximately the same for each quality level. Workers are largely satisfied with their responses (87\%~positive). These findings indicate that 
\begin{enumerate*}[label=(\arabic*)]
\item
workers have confidence in their work, despite low initial self-assessed expertise;
\item
the provided passages were considered sufficient;
\item
workers relied heavily on the passages to formulate their responses; and
\item
workers tend to curate the passages.
\end{enumerate*}

The reliability of this self-assessment, however, depends on worker faithfulness. We therefore conducted manual checks to judge overall response quality. Before hiring a worker, we provided an initial set of three topics. The submitted responses were manually screened and feedback was given to workers on assignment adherence and writing quality. If successful, a contract for additional batches of topics was drawn up. Our screening ensured that workers understand the assignment and drastically reduced spam.

To further ensure that responses were genuine human writing, we manually reviewed every response via a web interface that allows an accelerated replay of user interface interactions. We also watched out for common signs of LLM usage, such as copy-paste interactions of non-source material, sudden jumps in character count, completely homogeneous keyboard inputs, and absence of typing error corrections. This hybrid approach was trialed in a pilot study, where multiple instances of LLM usage were successfully spotted. Hired workers were made aware of our checks.

\subsubsection{Scale and Cost}

In total, 903~responses were collected; 86~responses were repeated due to workers failing our spam detection or screening for writing assistance tools. Some workers, who submitted responses that were rejected, were still compensated for their effort upon manual review, but all were discarded from the final response pool. A flat amount of~\$2.50 was paid per individual response, resulting in an hourly wage of~\$14.40 at an average work time of 10:22~minutes per response.
The total cost for collecting the human RAG responses was~\$2,610.32, which includes pilot study payments, platform fees, and tax.

\subsubsection{LLM Responses}
Using OpenAI's \texttt{GPT-4o} model, version 2024-08-06, we generated responses for each topic and discourse style, prompting it with the same instructions and passages given to the human workers.
The total inference cost was~\$23.32.

\subsection{Pairwise Utility Judgment}

With the second crowdsourcing study, we gather pairwise judgments for seven RAG utility dimensions. In line with previous work (Section~\ref{related-work}), we opt for a pairwise design for enhanced reliability, and verify this choice with a pointwise pilot study (Section~\ref{sub:pointwise-study}).

\subsubsection{Study Design}

We sampled~65 of the 301~TREC topics (60~topics as a 20\%~test split, 5~topics for pilot experiments). Each topic has six responses (three human-written, three LLM-generated). We thus add to the comparison pool per topic all 15~unique pairings of responses (975~pairs total) to be judged, and for a third of them (stratified by length and origin) the reverse pairing as well. The comparison pool comprises a total of 1,352~pairs. For each pair, utility is judged according to seven dimensions. Following \citet{gienapp:2024}, we adapt six of their ten dimensions, omitting external consistency and factual correctness, which require an in-depth comparative text analysis between RAG response and its sources, and both clarity dimensions, as they depend on the user information need, which has not been explicitly described for the TREC topics. Like \citet{hosking:2024}, we add an overall quality judgment as a baseline. All pairwise utility judgments for each of the utility dimensions permit a neutral response option, with the exception of overall quality, which mandates a definitive answer.

\begin{table}
    \small
    \caption{Pairwise operationalization of utility dimensions proposed by \citet{gienapp:2024}, as well as overall quality.}%
    \label{tab:utility-dimensions}
    \begin{tabularx}{\linewidth}{@{}lX@{}}
    \toprule
    \bfseries Dimension     & \textbf{Question} (Which response...)                                                \\ 
    \midrule 
    Topical Correctness     & ... better answers the query?                          \\
    Logical Coherence       & ... is easier to follow?                               \\
    Stylistic Coherence	    & ... has a more consistent writing style?               \\
    Broad Coverage          & ... looks at more aspects of the topic?                \\
    Deep Coverage           & ... explains things more in detail?                    \\
    Internal Consistency    & ... is clearer about how different views fit together? \\
    \midrule
    Overall Quality         & ... is better overall?                                 \\
    \bottomrule
    \end{tabularx}
\end{table}

A single questionnaire consists of 15~individual response pairs, each with its topic, randomly sampled and stratified by response length to ensure equal workload. We employ five different workers for each questionnaire. Workers receive preliminary instructions covering the study objective, interface guide, data presentation details, and judgment protocol.
We track user interaction data, e.g., clicks/dwell times for spam detection via the BigBro library~\citep{scells:2021}.

\subsubsection{Worker Recruitment}\label{sub:recruitment-pairwise}

We recruit workers via the Prolific platform,%
\footnote{\url{www.prolific.com}}
applying selection criteria to only recruit workers whose primary language is English, who live in countries with English as official language, who have an approval rate higher than~99\%, and who have previously completed at least 500~tasks.
In total, 420~individual workers were recruited. 
On average, a worker contributed judgments for 16~response pairs (minimum~11 / maximum~46).

\subsubsection{Review Process}

We identify three factors that could impact the quality of the collected judgments: 
\begin{enumerate*}[label=(\arabic*)]
\item
spam,
\item
misunderstood assignment, and
\item
under-performing workers.
\end{enumerate*}
We address spam by screening submissions by total time taken, with a lower acceptance limit of 10~minutes per questionnaire, i.e., 20~seconds per text in each response pair, corresponding to fast read time of 120~words (half of average text length). Submissions faster than that were rejected and re-judged by different workers. We address the other two factors through replication. Since response pairs are distributed randomly, i.e., a different set of workers judged each pair, the probability of the majority of workers simultaneously misunderstanding or under-performing is low per response pair. Further, we use MACE~\cite{hovy:2013} to estimate worker competency scores, inducing a trustworthy gold label for each pairwise comparison as a competence-weighted majority vote. Both these measures do not prevent systematic task failure, i.e., the majority of workers not being able to complete the task. We investigate this by cross-checking a sample of items with expert judgments (Section~\ref{sub:expert-study}), and find no evidence for systematic failure (Section~\ref{sub:reliability}).

\subsubsection{Scale and Cost}

In total, 47,320~unique pairwise judgments were collected, 6,760~for each of the seven utility dimensions. A flat amount of~\$5.50 was paid per individual questionnaire, resulting in an hourly wage of~\$13.20, with an average time of~25:00 minutes spent.
The total costs were~\$3,672.54, including pilot study costs, platform fees, and tax.

\subsubsection{LLM Judgments}\label{sec:llm-judgments}

We use OpenAI's \texttt{GPT-4o} model, version 2024-08-06, to collect pairwise utility judgments automatically. The model received the same instructions as given to the human judges, experimenting with two setups: judging all utility dimensions simultaneously for a given topic and response pair, and separately per dimension. We collect LLM judgments on all response pairs where humans judged both directions, for a total of 10,556~individual judgments. Inference was conducted with temperature zero to ensure consistent results. The total cost was~\$44.70.

\subsection{Verification Studies}

\subsubsection{Pointwise Judgments}\label{sub:pointwise-study}

As baseline for pairwise utility judgment, we also attempted to judge the seven utility dimensions in a pointwise manner. Each dimension is judged given 
\begin{enumerate*}
\item
a short question text, prompting for a judgment between two given extremes~A and~B of that utility dimension (e.g., `basic' and `detailed' for deep coverage); 
\item
a description, illustrating the judged concept and providing exemplary descriptions of both extremes; and
\item
a four-point Likert scale, (e.g., `very basic', `somewhat basic', `somewhat detailed', and `very detailed').
\end{enumerate*}
In total, 1,645~judgments were collected for a set of 47~unique, randomly chosen responses, each judged by five different workers. Workers were recruited and their work reviewed as described above (Section~\ref{sub:recruitment-pairwise}). A flat amount of~\$10.50 was paid per questionnaire, with an hourly wage of~\$13.40, for a total of~\$204.75, including platform fees and tax.

\subsubsection{Expert Judgments}\label{sub:expert-study}

We sample a set of 30~response pairs deemed hard to decide for crowd workers (see Section~\ref{sub:reliability}). Three persons from the author team served as expert judges. They used the same study design as the crowd workers.

\section{Analysis of RAG Response Writing}

\label{corpus-analysis-human-behavior}

Humans and LLMs reveal distinct patterns in their writing and use of documents when formulating responses. Our investigation into these patterns is structured around five key questions:
\begin{enumerate*}[label=(\arabic*)]
    \item How many and which of the documents are used by workers? 
    \item Does the given ranking order matter, or do workers treat the retrieved documents as a set? 
    \item Is information transferred verbatim, or paraphrased? 
    \item Where and how are citations located within the text? 
    \item How accessible is the final response to readers?
\end{enumerate*} 

This section thus presents a systematic examination of response characteristics through descriptive statistics. In congruence with the posed questions, it analyses document selection, citation order, text reuse, attribution position and granularity, and readability, for both human workers and LLMs. We segment the raw text responses into individual statements using the explicit citation markers as delimiters, with both human workers and LLMs having received identical citation formatting instructions (bracketed numbers). 

\subsection{General Characteristics}

\begin{figure}
    \includegraphics[width=\linewidth]{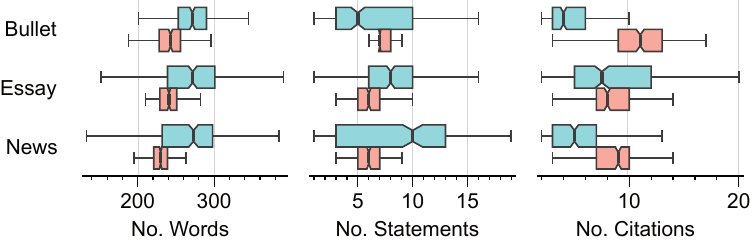}
    \Description{A plot consisting of three sub-figures. Each one shows boxplots of values, separated in lines by text style, i.e., bullet, essay, and news, and separated in color for origin, i.e., LLM-generated or human-written. The first boxplots graphs number of words per response, the second graphs number of statements per response, the third plots number unique citations per response.}
    \caption{Distribution of space-separated words, citation-separated statements, and unique cited documents, per type and origin (\raisebox{1.5pt}{\fcolorbox{Red2}{Red5}{\rule{0pt}{2pt}\rule{2pt}{0pt}}} LLM, \raisebox{1.5pt}{\fcolorbox{Blue2}{Blue5}{\rule{0pt}{2pt}\rule{2pt}{0pt}}} Human).}
    \label{fig:descriptive-statistics}
\end{figure}

Figure~\ref{fig:descriptive-statistics} presents the distribution of word count, statement count, and citation count over responses across writing styles and origins, excluding non-responses below 50 words. Both human workers and LLMs remain close to the 250-word target, with LLM responses exhibiting lower variance and marginally shorter lengths. They also contain fewer statements with reduced variation in count, yet employ more unique documents: humans use an average of 6.3 documents of the available 20 per response, while LLMs draw from 9.2. Moreover, human workers cite more individual documents (2.1 citations per document versus LLMs' 1.35) but incorporate fewer documents per statement (1.62 versus LLMs' 1.93). This pattern, combined with the humans' shorter statement length, indicates more granular information attribution: humans segment information into smaller units with precise document attribution, while LLMs favor broader statements, synthesizing multiple documents.

\subsection{Document Selection}

\begin{table}
    \small
    \caption{Mean and 95\% CI of Jaccard coefficient and Spearmans' $\rho$ correlation, for different pairings of citation sets.}
    \label{tab:citation-measures}
    \renewcommand{\tabcolsep}{4pt}
    \begin{tabularx}{\linewidth}{@{}Xcc @{}S[table-format=-2.2+-1.2] S[table-format=-1.2+-1.2] S[table-format=-1.2+-1.2] S[table-format=1.2]@{}}
    \toprule 
    Measure     & \multicolumn{2}{@{}c@{}}{Pair} & {Bullet}   & {Essay}   & {News}    & {Avg.} \\
    \midrule            
    \multirow{3}{*}{Jaccard-Coeff.\kern-1em}     
            & \emojihuman & \emojiranking         & 0.22(012) & 0.43(025) & 0.29(022) & 0.31 \\
            & \emojimodel & \emojiranking         & 0.54(014) & 0.41(011) & 0.43(011) & 0.46 \\
            & \emojihuman & \emojimodel           & 0.30(017) & 0.38(018) & 0.31(016) & 0.33 \\
    \midrule
    \multirow{3}{*}{Spearmans' $\rho$\kern-1em}  
            & \emojihuman & \emojiranking         & 0.51(061) & 0.35(052) & 0.33(064) & 0.39 \\
            & \emojimodel & \emojiranking         & 0.24(039) & 0.34(044) & 0.30(048) & 0.29 \\
            & \emojihuman & \emojimodel           & 0.39(070) & 0.36(062) & 0.32(071) & 0.36 \\
    \bottomrule
    \\[-2ex]
    \multicolumn{7}{c}{\emojihuman = Human, \emojimodel = LLM, \emojiranking = Document Ranking}
    \end{tabularx}
\end{table}

\begin{figure}
\small
\begin{minipage}{.5\linewidth}
    \includegraphics[width=\linewidth]{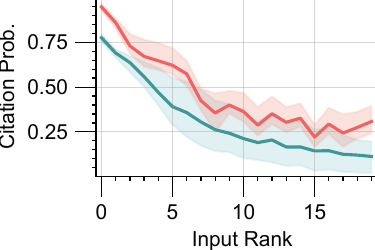}
\end{minipage}\hspace{.04\linewidth}
\begin{minipage}{.45\linewidth}
    \begin{tabular}{@{}l*{2}{S[table-format=1.1]}*{2}{S[table-format=2.1]}@{}}
\toprule
Style   & \multicolumn{2}{c}{Cited} & \multicolumn{2}{c}{Uncited} \\
        \cmidrule(lr){2-3} \cmidrule(l){4-5}
        & {\emojihuman} & {\emojimodel} & {\emojihuman} & {\emojimodel} \\
\midrule
Bullet  & 3.7 & 7.2 & 11.2 & 11.8 \\
Essay   & 6.5 & 6.3 & 11.6 & 11.5 \\
News    & 4.7 & 6.6 & 11.1 & 11.5 \\
\midrule
Avg.    & 5.0 & 6.7 & 11.3 & 11.6 \\
\bottomrule
\end{tabular}

\end{minipage}%
\Description{A plot showing the probability of a source being cited on the x-axis, and the position of the source in the input ranking on the x-axis. Two lines are graphed: one for LLM-generated responses, one for Human-written texts. Both lines feature areas of uncertainty calculated by aggregating over all responses of each kind.}
\caption{Left: probability of a document being cited by rank position; Right: median rank of cited and uncited documents, per response style and origin (\emojimodel/\raisebox{1.5pt}{\fcolorbox{Red2}{Red5}{\rule{0pt}{2pt}\rule{2pt}{0pt}}} LLM, \emojihuman/\raisebox{1.5pt}{\fcolorbox{Blue2}{Blue5}{\rule{0pt}{2pt}\rule{2pt}{0pt}}} Human).}
\label{fig:median-ranking-selection}
\end{figure}

To more closely examine document selection criteria, we calculate the Jaccard coefficient between cited and available documents. In an extension of the overall citation counts as shown in Figure~\ref{fig:descriptive-statistics}, the first partition of Table~\ref{tab:citation-measures} breaks down document selection by text style. The disparity in document count is most pronounced in bullet-style responses, where humans cite their minimum of 4.4 documents against LLMs' maximum of 10.8. For essay-style responses, almost equal counts of documents are used, while news-style responses again exhibit a slightly higher count for LLM-written responses. The specific documents used differ, too, with a Jaccard overlap of only 0.33 on average between human-cited and LLM-cited documents. This selection in both cases appears to be influenced by the order of documents. In Figure~\ref{fig:median-ranking-selection} (left), the probability of a document being cited decreases with ranking position. In Figure~\ref{fig:median-ranking-selection} (right), the table shows this effect across all writing styles, as the median rank (averaged across all topics) of documents cited in the response is much lower, i.e., closer to the top, than for uncited documents.

\subsection{Citation Order} 

While rank position influences document selection, it might also influence the structure of the text, evident if the input ranking order is similar to the documents' citation order in-text. We calculate Spearman's $\rho$ correlation coefficient between a documents' rank position and its relative position in the response, excluding uncited documents. The second partition of Table~\ref{tab:citation-measures} details the resulting distribution of $\rho$. Both human workers and LLMs demonstrate weak positive correlations between citation and ranking order (first two lines), with humans exhibiting consistently higher correlations across all response styles. Bullet-style responses show the strongest adherence to initial ranking, while news-style responses score lowest. Human responses display greater variance in correlation values suggesting that they make more pronounced deviations when choosing alternative orderings. However, an average correlation of only 0.39 for humans and 0.29 for LLMs indicates that both primarily treat documents as unordered sets. When examining the overlapping subset of documents cited by both groups (last line), the low correlation (0.37) indicates that both re-rank differently. 

\subsection{Text Reuse}

\begin{figure}
    \includegraphics[width=\linewidth]{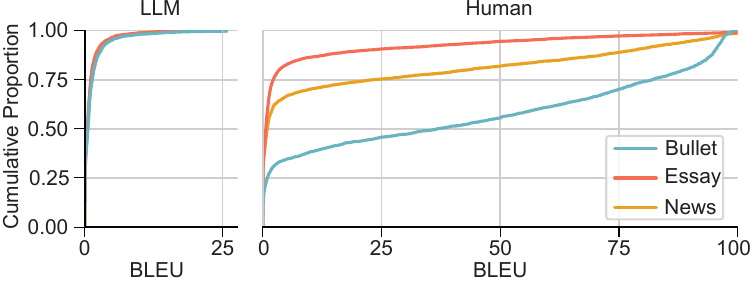}
    \caption{Cumulative proportion of statement/reference pairs per sentence-BLEU score.}
    \Description{A plot showing 2 graphs; both show the cumulative proportion of all texts on the y-axis between 0 and 1, and the BLEU score from 0 to 100 on the x-axis. The left plot shows data for LLM-generated texts, the right plot shows data for Human-written texts. Each plot contains 3 separate curves, one for each of the text styles Bullet, Essay, and News.}
    \label{fig:citations-bleu-scores}
\end{figure}

To quantify the occurrence of text reuse, i.e., copying between documents and response, we compute sentence-BLEU~\cite{post:2018} between each statement of a response and its cited documents, using a maximum ngram-order of 8. Figure~\ref{fig:citations-bleu-scores} displays the cumulative score distributions per response style and origin. LLM responses consistently show sentence-BLEU scores below 25 with no deviation across styles, while human responses exhibit style-dependent variation. For news and essay styles, 75\% and 88\% of human responses respectively score below the maximum LLM value of 25. Bullet-style responses demonstrate substantially higher verbatim copying. This style-specific divergence reflects distinct writing strategies. A cursory qualitative analysis of bullet-style responses reveals that LLMs primarily organize responses by answer aspects, each citing multiple documents and building an abstractive micro-summary around it, while humans tend to organize responses by document rank position, favoring passage reuse and creating bullet lists that mirror list-SERP anchor texts.

\subsection{Attribution Position \& Granularity}

\begin{figure}
    \includegraphics[width=\linewidth]{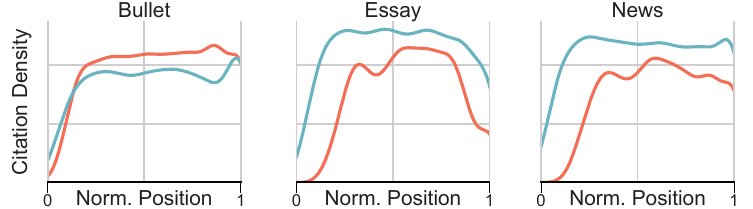}
    \caption{Relative density of citations over normalized text position per style and origin (\raisebox{1.5pt}{\fcolorbox{Red2}{Red5}{\rule{0pt}{2pt}\rule{2pt}{0pt}}} LLM, \raisebox{1.5pt}{\fcolorbox{Blue2}{Blue5}{\rule{0pt}{2pt}\rule{2pt}{0pt}}} Human).}
    \Description{A plot showing 3 graphs, one for each of the three writing styles Bullet, Essay, and News. The y-axis represents the relative citation density, the x-axis represents normalized text position, between 0 and 1. Each of the three plots shows 2 separate curves, one for LLM, one for Human texts.}
    \label{fig:citations-density}
\end{figure}

Figure~\ref{fig:citations-density} illustrates the relative citation density across normalized text positions, revealing distinct patterns in reference distribution. All response styles exhibit an initial lag, as there is always text preceding the first citation, followed by consistent citation density throughout the text. This onset is more pronounced for LLM-written responses, indicating that their first citation occurs later on average. Additionally, essay-style responses uniquely show decreased citation density in their concluding segments, suggesting unreferenced summary statements. Humans demonstrate higher overall citation density in essay and news styles, indicating more granular source attribution. Bullet-style responses, however, show comparable citation densities between humans and LLMs.

\subsection{Readability}

\begin{figure}
    \includegraphics[width=\linewidth]{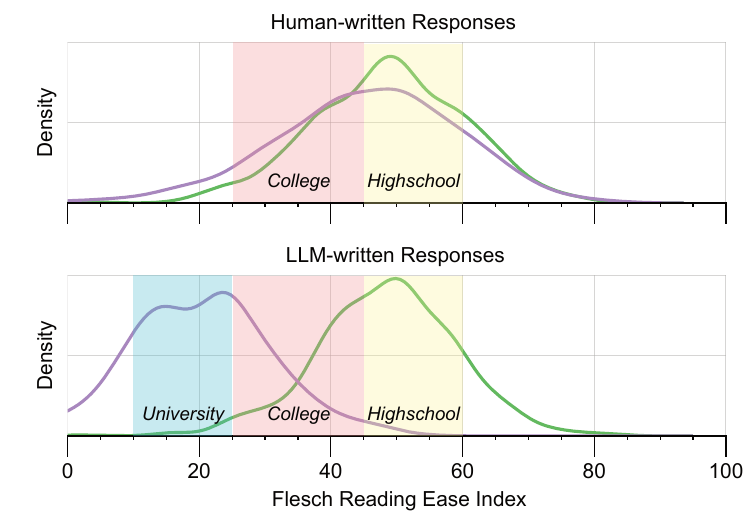}
    \Description{A plot showing 2 graphs of Flesch reading ease index scores. The upper one plots the density of scores across all Human-written responses, the lower one plots the same but for LLM-written responses. Both plots feature overlays for the score ranges corresponding to university, college, and high school reading levels.}
    \caption{Distribution of Flesch reading ease index scores for responses \raisebox{1.5pt}{\fcolorbox{Purple2}{Purple5}{\rule{0pt}{2pt}\rule{2pt}{0pt}}} and their cited documents \raisebox{1.5pt}{\fcolorbox{Green2}{Green5}{\rule{0pt}{2pt}\rule{2pt}{0pt}}}, per origin.}
    \label{fig:readability}
\end{figure}

To judge and compare the accessibility of responses, we compare text readability scores between human and LLM-generated content, as well as the documents they cite. We calculate Flesch reading ease indices~\cite{kincaid:1975} at the statement level, for both the statement text and the document it cites. An aggregate score pair for each response and its sources is built using the median of its respective statements' and the median of the cited documents' scores. Figure~\ref{fig:readability} plots the distribution of both readability scores for human- and LLM-written texts, respectively. Negligible deviation was apparent between text styles in both cases, hence their distinction is omitted from the plot.
Human-authored responses maintain readability levels matching their cited documents, both exhibiting normally distributed scores centered around a score of 50, corresponding to high school reading level. In contrast, while the cited documents used by LLMs follow a similar normal distribution centered around a score of 50, their generated text shows markedly lower readability with a distribution centered at slightly above 20, corresponding to college- to university-level reading ease. This suggest that LLMs tend to increase text complexity when reformulating content, potentially due to their tendency to combine aspects from multiple documents into densely packed statements, while humans, as shown in previous subsections, align their writing more closely to the cited documents.

\subsection{Summary}

Human and LLM writing behaviors differ in several aspects. LLMs produce responses with lower variance in length and fewer statements, but draw from a broader set of documents. Document selection differs, though both LLMs and human workers are influenced by input ranking order. Citation sequencing demonstrates weak correlations with input order, indicating treatment of the top documents as unordered sets. Text reuse analysis shows LLMs exhibit low source overlap, while humans tend to include more verbatim short-span reuse. Readability judgment demonstrates humans maintain reading ease, while LLMs generate more complex text.

\section{Crowdsourcing for RAG Evaluation}

\label{rag-response-preferences}

We investigate the suitability of crowdsourcing to gather data for both reference-based and judgment-based RAG~evaluation guided by four central questions:
\begin{enumerate*}[label=(\arabic*)]
    \item Is crowdsourcing suitable for utility judgments?
    \item How do human-written and LLM-generated RAG responses compare w.r.t. different utility dimensions in judgment-based evaluation?
    \item Do reference-based evaluation metrics succeed in reproducing the system ranking given by preference data?
    \item Can LLMs successfully serve as utility judges, reproducing crowd judgment?
\end{enumerate*}
This section thus present a systematic examination of the reliability of crowdsourced data, the two RAG evaluation paradigms, and the competing LLM-as-judge approach.

\subsection{Crowdsourcing Reliability}\label{sub:reliability}

We establish the reliability of the collected data by investigating agreement, presence of order bias, and utility interdependence. 

\paragraph{Agreement \& Competency}

\begin{table}[t]
\small 
\renewcommand{\arraystretch}{.8}
\caption{Krippendorff's $\alpha$ agreement for pairwise RAG utility judgment. The minimal differentiability split includes items where for a majority of dimensions, no majority vote is established by the five crowd judges. `Pntw' denotes pointwise judgments from the verification study, $n_\text{Items} = 410$.}
\label{tab:agreement}
\begin{tabularx}{\linewidth}{@{}X *{7}{S[table-format=-1.2,table-column-width=.7cm]}@{}}
\toprule
Util. D.        & \multicolumn{2}{c}{Complete Data}                 & \multicolumn{2}{c}{Min.-Diff. Split}          & \multicolumn{2}{c@{}}{Diff. Split}            & \multicolumn{1}{c}{\multirow{2}{*}{\rotatebox[origin=c]{90}{Pntw.}}}\\
                    & \multicolumn{2}{c@{}}{$n_{\text{Items}} = 1352$}  & \multicolumn{2}{c}{$n_{\text{Items}} = 290$}  & \multicolumn{2}{c}{$n_{\text{Items}} = 1062$} &\\
                    \cmidrule(lr){2-3}                                  \cmidrule(lr){4-5}                              \cmidrule(l){6-7}
                    & {$\alpha$ (C)}        & {$\alpha$ (C*)}           & {$\alpha$ (C)}        & {$\alpha$ (E)}        & {$\alpha$ (C)}    & {$\alpha$ (C*)}           & {$\alpha$ (C*)}\\
\midrule
Top. Cor.        & 0.19                  &  0.43                     & -0.11                 & -0.01                 & 0.28              & 0.48                      & 0.22 \\
Log. Coh.        & 0.18                  &  0.39                     &  0.03                 &  0.09                 & 0.23              & 0.46                      & 0.21 \\
Styl. Coh.       & 0.11                  &  0.38                     & -0.08                 &  0.08                 & 0.17              & 0.44                      & 0.21 \\
Brd. Cov.        & 0.28                  &  0.44                     & -0.07                 &  0.01                 & 0.37              & 0.51                      & 0.23 \\
Dp. Cov.         & 0.28                  &  0.45                     & -0.04                 &  0.10                 & 0.36              & 0.51                      & 0.21 \\
Int. Con.        & 0.14                  &  0.42                     & -0.12                 & -0.14                 & 0.22              & 0.49                      & 0.21 \\
Ovr. Qu.         & 0.17                  &  0.39                     & -0.12                 &  0.08                 & 0.24              & 0.47                      & 0.19 \\
\midrule
Mean             & 0.19                  & 0.41                      & -0.07                 &  0.03                 & 0.27              & \textbf{0.48}             & 0.21 \\      
\bottomrule
\\[-2ex]
\multicolumn{7}{c}{\footnotesize E = Expert, C = Crowd, C* = Competency-corrected Crowd}
\end{tabularx}
\end{table}

We assess crowd judgments reliability using Krippendorff's $\alpha$ for ordinal judgments. Table~\ref{tab:agreement} shows agreement values for the complete set of collected judgments (C). Initial agreement appears low, averaging 0.19 with substantial variance across utility dimensions. Yet, two primary sources of low agreement emerge: worker competency limitations and pairs with minimal differentiability, where both responses are too similar to enable consistent comparative judgments. To address these challenges, we employ two complementary strategies. First, we use MACE~\cite{hovy:2013} to estimate worker competency scores and generate competency-corrected gold labels. Judgments submitted by the lowest-scoring workers in terms of competency were subsequently removed from the judgment pool, ensuring that at least three judgments remain for all pairs. This excludes approximately the lower quarter of the workers from consideration (C*). This substantially improves agreement, more than doubling the average to 0.41 with reduced variance. Second, we identify minimally differentiable item pairs through voting behavior: pairs lacking a majority vote (3 out of 5) across a majority of utility dimensions (4 out of 7). Approximately 20\% of pairs fall into this category. A targeted expert validation of a 30-item sub-sample reveals that while experts demonstrate higher absolute agreement, both crowd and expert workers fail to establish a reliable vote. Notably, experts show a higher tendency for neutral choices, which crowd workers tend to avoid, thereby amplifying disagreement. We thus conclude that agreement is negatively impacted by both a small set of low-competency crowd workers and a small set of minimally differentiable items. After applying correction for both factors, the resulting gold labels are subject to agreement levels (marked bold in Table) comparable and even exceeding previous IR and NLP literature~\cite{zhang:2024,goyal:2022} and we deem them reliable. 
Additionally, investigate the agreement attainable in a pointwise study design (Section~\ref{sub:pointwise-study}) for comparison. After applying the same MACE competency correction, the resulting agreement is still very low at \SI{0.21}{}, not better than the uncorrected pairwise scores. The pairwise study design, while more expensive, thus offers a substantial improvement in reliability. More rigorous worker competency screening could help improve cost efficiency.  

\paragraph{Order Bias}

\begin{table}[t]
\small 
\renewcommand{\arraystretch}{.8}
\caption{Krippendoff's $\alpha$ agreement and maximum delta between individual directions ($\leftarrow$/$\rightarrow$) and combined set ($\leftrightarrow$) for within-item order effects, and $1^{\text{st}}$ and $2^{\text{nd}}$ half of items in a questionnaire for across-item order effects.}\label{tab:order-effect}
\begin{tabularx}{\linewidth}{@{}X *{4}{S[table-format=1.2]} *{3}{S[table-format=1.2,table-column-width=.79cm]}@{}}
\toprule
Util. D.      & \multicolumn{4}{c}{Within-item Order} & \multicolumn{3}{c@{}}{Across-item Order} \\
                & \multicolumn{4}{c}{$n_{\text{Items}} = 377$} & \multicolumn{3}{c@{}}{$n_{\text{Items}} = 1352$}\\
                    \cmidrule(lr){2-5} \cmidrule(l){6-8}
                & {$\alpha (\leftarrow$)} & {$\alpha (\rightarrow$)} & {$\alpha (\leftrightarrow$)} & {$\Delta_\text{max}$} & {$\alpha (1^{\text{st}})$} & {$\alpha (2^{\text{nd}})$} & {$\Delta$} \\
\midrule
Top. Cor.                & 0.18 & 0.19 & 0.17 & 0.02 & 0.18 & 0.20 & 0.01 \\
Log. Coh.\kern-1em       & 0.18 & 0.19 & 0.18 & 0.01 & 0.18 & 0.18 & 0.00 \\
Styl. Coh.\kern-1em     & 0.08 & 0.14 & 0.10 & 0.04 & 0.11 & 0.11 & 0.00 \\
Brd. Cov.                & 0.27 & 0.28 & 0.26 & 0.02 & 0.28 & 0.28 & 0.00 \\
Dp. Cov.                 & 0.27 & 0.27 & 0.26 & 0.01 & 0.28 & 0.27 & 0.01 \\
Int. Con.\kern-1em       & 0.13 & 0.15 & 0.14 & 0.01 & 0.13 & 0.16 & 0.02 \\
Ovr. Qu.                 & 0.16 & 0.17 & 0.15 & 0.02 & 0.15 & 0.18 & 0.03 \\
\midrule
Mean                     & 0.18 & 0.20 & 0.18 & 0.01 & 0.19 & 0.20 & 0.01 \\
\bottomrule
\end{tabularx}
\end{table}

We distinguish two types of order effects: within-item and across-item. Within-item order effects occur when the presentation sequence of two responses within a single questionnaire item influences judge decisions, such as a preference for the first option. Across-item order effects occur when the sequence of items in a questionnaire systematically impacts judgments, potentially due to learning or fatigue. As presented in Table~\ref{tab:order-effect}, for within-item order, we examine the subset of comparisons where both response directions were independently judged. Under order bias, we would expect higher agreement among workers when analyzing each direction ($\leftarrow/\rightarrow$) separately compared to pooled results ($\leftrightarrow$). For across-item order, we split the dataset by item position in each worker's questionnaire, i.e., the first seven response pairs ($1^{\text{st}}$) and the remaining response pairs ($2^{\text{nd}}$) of each questionnaire. As questionnaire order is randomized, any difference in agreement between the two halves would originate from study design, not data characteristics. For both effect types, the agreement values remain nearly identical, both overall as well as for individual utility dimensions, suggesting robustness to presentation order. 

\paragraph{Interdependence of Utility}

\begin{table}[t]
\small
\renewcommand{\arraystretch}{.8}
\caption{Kendall's $\tau$ correlation cross-tabulation between gold labels of all quality dimensions. Columns use initial letters to designate utility dimensions.}
\label{tab:cross-correlation}
\begin{tabularx}{\linewidth}{@{}X*{7}{S[table-format=1.2]}@{}}
\toprule
Utility Dimension       & {T}  & {L}   & {S}   & {B}   & {D}   & {I}   & {O}   \\
\midrule
{T}opical Correctness   &      &  0.30 &  0.35 &  0.42 &  0.46 &  0.45 &  0.52 \\
{L}ogical Coherence     & 0.30 &       &  0.34 &  0.18 &  0.20 &  0.30 &  0.37 \\
{S}tylistic Coherence   & 0.35 &  0.34 &       &  0.23 &  0.29 &  0.36 &  0.36 \\
{B}road Coverage        & 0.42 &  0.18 &  0.23 &       &  0.57 &  0.38 &  0.42 \\
{D}eep Coverage         & 0.46 &  0.20 &  0.29 &  0.57 &       &  0.40 &  0.43 \\
{I}nternal Consinstency & 0.45 &  0.30 &  0.36 &  0.38 &  0.40 &       &  0.45 \\
{O}verall Quality       & 0.52 &  0.37 &  0.36 &  0.42 &  0.43 &  0.45 &       \\ 
\midrule
Mean                    & 0.42 &  0.28 &  0.32 &  0.37 &  0.39 &  0.39 &  0.43 \\
\bottomrule
\end{tabularx}
\end{table}

To assess crowd workers' ability to discriminate utility dimensions, we investigate interdependence by computing Kendall's $\tau$ correlation between gold labels across dimensions as shown in Table~\ref{tab:cross-correlation}. Values range from 0.18 to 0.57, with a mean of 0.37. While overall quality exhibits the highest average correlation (0.42), the low inter-dimensional correlations of others suggests successful differentiation. We further find indications of thematic clustering: coverage dimensions and topical correctness -- addressing response groundedness -- demonstrate higher inter-correlation while exhibiting distinctly low correlation with stylistic and logical coherence, addressing response presentation.

\subsection{Judgment-based Evaluation}

With the reliability of the preference data established, we operationalize the judgment-based evaluation approach, ranking the given responses per topic based on the pairwise comparisons, in order to comparatively quantify the utility of human-written and LLM-generated RAG responses. To rank the six different responses within each topic based on the pairwise preferences, we use a variant of the probabilistic Bradley-Terry model adapted to handle ties and contradictory comparisons~\cite{gienapp:2020}. It has been successfully applied to infer robust scalar scores from crowdsourced pairwise labels on text quality before~\cite{gienapp:2020}. We compute ranks independently for each topic and utility dimension, and assign based on them descending `grades' between 6 and 1 (higher is better). This allows us to compare grades across topics. Table~\ref{tab:pairwise-grading} shows the mean grade for each combination of text style, origin, and utility dimension. Further, it includes averages per style without distinguishing origin (columns `Both'), per origin without distinguishing style (column `All Styles'), as well as across all utility dimensions (last row). 

LLMs consistently received higher grades than human workers across most evaluation dimensions. This difference is statistically significant (Wilcoxon signed rank test, $\alpha = 0.05$, Benjamini-Hochberg correction) for most style-dimension combinations, with two notable exceptions: logical coherence in essay-style responses, both coverage dimensions (broad and deep) in news-style responses, and deep coverage for bullet-style responses. When aggregating across styles, LLMs higher grading remained significant for all utility dimensions except coverage. In a style-based analysis independent of authorship, bullet-style responses significantly outperformed both essay and news styles, which are graded on-par (Wilcoxon signed rank test, $\alpha < 0.05$, Benjamini-Hochberg correction). Bullet-style responses showed particular strength in logical coherence, coverage metrics, and overall quality, while stylistic coherence and internal consistency showed no significant style-based differences. Between essay and news styles, the only significant difference emerged in deep coverage, favoring essays.

\begin{table}
\renewcommand{\emojiboth}{\kern-1.05pt\raisebox{.25ex}[0pt][0pt]{\emojihuman}\kern-.5em\raisebox{-.25ex}[0pt][0pt]{\emojimodel\kern-1.05pt}}

\caption{Mean response grade per style, origin, and utility dimension. \emojiboth~designates union of both origins. Grade is on a scale of 1--6, higher is better.}
\label{tab:pairwise-grading}
\small
\renewcommand{\arraystretch}{.8}
\begin{tabularx}{\linewidth}{
    @{}X
    *{11}{S[table-format=1.1,table-column-width=.35cm]}
    @{}
}
\toprule
{Util.}      &\multicolumn{3}{c}{Bullet}                     &\multicolumn{3}{c}{Essay}                     &\multicolumn{3}{c}{News}                       &\multicolumn{2}{c}{All Styles}       \\
                \cmidrule(r){2-4}                               \cmidrule(lr){5-7}                             \cmidrule(lr){8-10}                              \cmidrule(l){11-12}
                & {\emojihuman} & {\emojimodel} & {\emojiboth}        & {\emojihuman} & {\emojimodel} & {\emojiboth}       & {\emojihuman} & {\emojimodel} & {\emojiboth}        & {\emojihuman} & {\emojimodel} \\
\midrule

{Top.} & 2.83(146)    & 4.54(138)    & 3.68(166)    & 3.06(169)    & 4.14(167)    & 3.60(176)   & 2.65(172)    & 3.78(146)    & 3.22(169)    & 2.85(163)    & 4.15(153)    \\
{Log.} & 4.22(145)    & 5.38(103)    & 4.80(138)    & 2.71(164)    & 3.08(137)    & 2.89(152)   & 2.35(133)    & 3.26(140)    & 2.81(143)    & 3.09(168)    & 3.91(165)    \\
{Sty.} & 3.20(149)    & 4.25(160)    & 3.72(163)    & 2.57(139)    & 4.17(158)    & 3.37(169)   & 2.58(155)    & 4.23(167)    & 3.41(180)    & 2.78(150)    & 4.22(161)    \\
{Bro.} & 3.52(167)    & 4.55(154)    & 4.04(168)    & 2.89(182)    & 3.78(148)    & 3.34(171)   & 2.83(173)    & 3.42(143)    & 3.12(161)    & 3.08(176)    & 3.92(155)    \\
{Dee.} & 3.58(182)    & 4.25(151)    & 3.92(170)    & 3.06(169)    & 4.11(151)    & 3.58(168)   & 2.72(176)    & 3.28(145)    & 3.00(163)    & 3.12(179)    & 3.88(154)    \\
{Int.} & 3.02(165)    & 4.26(142)    & 3.64(166)    & 2.94(182)    & 4.22(137)    & 3.58(173)   & 2.63(163)    & 3.94(160)    & 3.28(174)    & 2.86(170)    & 4.14(147)    \\
{Ovr.} & 3.28(168)    & 4.89(124)    & 4.08(168)    & 2.71(167)    & 4.00(150)    & 3.35(171)   & 2.54(154)    & 3.58(147)    & 3.06(159)    & 2.84(165)    & 4.16(150)    \\
       \cmidrule(r){2-4}                               \cmidrule(lr){5-7}                             \cmidrule(lr){8-10}                              \cmidrule(l){11-12}
{Avg.} & 3.38(160)    & 4.59(139)    & 3.89(163)    & 2.85(167)    & 3.93(150)    & 3.39(169)   & 2.62(161)    & 3.64(150)    & 3.13(164)    & 2.95(167)    & 4.05(155)    \\

\bottomrule
\end{tabularx}
\end{table}

\subsection{Reference-based Evaluation}

\begin{table}
\small
\renewcommand{\arraystretch}{.8}
\caption{Spearman rank correlation between label-induced ranking and ranking by content overlap metrics. Best/Worst only compares relative order of highest and lowest-ranked according to label ranking.}
\label{tab:content-overlap-metrics}
\begin{tabularx}{\linewidth}{@{}X*{6}{S[table-format=1.3]}@{}}
\toprule
Util. Dim.      & \multicolumn{3}{c}{Best/Worst} & \multicolumn{3}{c}{Full Ranking} \\
                \cmidrule(lr){2-4} \cmidrule(l){5-7}
                &{BERTS.}& {BLEU} & {RougeL} & {BERTS.} & {BLEU} & {RougeL} \\
\midrule
{T}opical Cor   & 0.354 & 0.354 & 0.169 & 0.172 & 0.211 & 0.094 \\
{L}ogical Coh.  & 0.354 & 0.477 & 0.446 & 0.288 & 0.337 & 0.294 \\
{S}tylistic Coh.& 0.385 & 0.385 & 0.385 & 0.263 & 0.315 & 0.243 \\
{B}road Cov.    & 0.231 & 0.354 & 0.231 & 0.183 & 0.254 & 0.228 \\
{D}eep Cov.     & 0.262 & 0.415 & 0.385 & 0.211 & 0.269 & 0.223 \\
{I}nternal Con. & 0.323 & 0.323 & 0.323 & 0.191 & 0.208 & 0.180 \\
{O}verall Qu.   & 0.385 & 0.323 & 0.292 & 0.292 & 0.268 & 0.255 \\
\bottomrule
\end{tabularx}
\end{table}

To investigate the efficacy of content-overlap and similarity metrics for RAG~evaluation, we leverage the pairwise-judgment-based response rankings established per topic previously. For each topic, we designated the highest-ranked response as the reference and generated rankings for the remaining five candidate responses. We use three metrics to rank candidates with respect to the reference: the contextualized-embedding-based BERTScore~\cite{zhang:2020}, and two ngram overlap measures - sentence-BLEU~\cite{post:2018} and RougeL~\cite{lin:2004}. We then compute Spearman's rank correlation between these metric-induced rankings and the ground-truth rankings derived from human judgments. This approach tests the hypothesis that effective re\-fe\-ren\-ce-based evaluation produces rankings consistent with human judgment. Table~\ref{tab:content-overlap-metrics} presents correlation values across utility dimensions, examining both binary discrimination (correct ordering of best and worst candidate according to ground-truth ranking) and full ranking correlation. Across all metrics and utility dimensions, correlation remains low. The best/worst correlation is slightly better than full ranking correlation, indicating that higher differentiability of responses within the respective quality dimension corresponds to better efficacy of reference-based evaluation; yet, fine-grained accurate system rankings, as in the full correlation setup, remain problematic. BLEU shows highest correlation for all utility dimensions except overall quality, where BERTScore performs best. The highest correlation in all metrics and both is attained for logical coherence. Yet, given an absolute maximum value of merely \SI{0.477}{} (BLEU for logical coherence), none of the  utility dimensions are accurately measurable by any of the three metrics. 

\subsection{LLMs as Judges of Utility}

\begin{table}
\renewcommand{\arraystretch}{.8}
\small
\setlength{\tabcolsep}{3pt}
\caption{Krippendorff $\alpha$ agreement between LLM prompt variations ($\alpha$, {\raisebox{.5ex}[0pt][0pt]{\emojimodel}\kern-.5pt/\kern-.5pt\raisebox{-.5ex}[0pt][0pt]{\emojimodel}}) / LLM and human gold judgment ($\alpha$, {\raisebox{.5ex}[0pt][0pt]{\emojimodel}\kern-.5pt/\kern-.5pt\raisebox{-.5ex}[0pt][0pt]{\emojihuman}}), and mean judgment correlation to other dimensions ($\bar{\rho}$, \emojimodel) for  different prompting strategies by utility dimensions.}
\label{tab:llm-as-judge}
\begin{tabularx}{\linewidth}{@{}X*{7}{S[table-format=1.2]}@{}}
\toprule
Utility Dimension       & \multicolumn{3}{c}{Combined Inference} & \multicolumn{3}{c}{Individual Inference} & Both\\
                        \cmidrule(lr){2-4} \cmidrule(lr){5-7}
                        & \multicolumn{2}{c}{$\alpha$} & {$\bar{\rho}$} & \multicolumn{2}{c}{$\alpha$} & {$\bar{\rho}$} & {$\alpha$}\\
                        \cmidrule(lr){2-3}                              \cmidrule(lr){5-6}                                  
                        & {\raisebox{.5ex}[0pt][0pt]{\emojimodel}\kern-.5pt/\kern-.5pt\raisebox{-.5ex}[0pt][0pt]{\emojimodel}} & {\raisebox{.5ex}[0pt][0pt]{\emojimodel}\kern-.5pt/\kern-.5pt\raisebox{-.5ex}[0pt][0pt]{\emojihuman}} & {\emojimodel} &{\raisebox{.5ex}[0pt][0pt]{\emojimodel}\kern-.5pt/\kern-.5pt\raisebox{-.5ex}[0pt][0pt]{\emojimodel}} & {\raisebox{.5ex}[0pt][0pt]{\emojimodel}\kern-.5pt/\kern-.5pt\raisebox{-.5ex}[0pt][0pt]{\emojihuman}} & {\emojimodel} & {\raisebox{.5ex}[0pt][0pt]{\emojimodel}\kern-.5pt/\kern-.5pt\raisebox{-.5ex}[0pt][0pt]{\emojimodel}}     \\
                        \midrule                
{T}opical Correctness   &  0.65 & 0.10 &  0.80 &  0.87 &  0.12 &  0.86 & 0.48 \\
{L}ogical Coherence     &  0.78 & 0.16 &  0.74 &  0.84 &  0.16 &  0.82 & 0.53 \\
{S}tylistic Coherence   &  0.82 & 0.14 &  0.71 &  0.88 &  0.13 &  0.82 & 0.53 \\
{B}road Coverage        &  0.65 & 0.08 &  0.64 &  0.58 &  0.06 &  0.74 & 0.40 \\
{D}eep Coverage         &  0.38 & 0.01 &  0.72 &  0.37 &  0.00 &  0.74 & 0.32 \\
{I}nternal Consinstency &  0.66 & 0.12 &  0.77 &  0.90 &  0.13 &  0.84 & 0.49 \\
{O}verall Quality       &  0.64 & 0.06 &  0.81 &  0.86 &  0.09 &  0.86 & 0.47 \\
\bottomrule
\end{tabularx}
\end{table}

To investigate the ability of LLMs as utility judges, we assess three critical properties:
\begin{enumerate*}[label=(\arabic*)]
\item consistency in judgment across conditions,
\item correctness relative to human gold labels, and
\item dimensional differentiation in utility judgment.
\end{enumerate*}
Table~\ref{tab:llm-as-judge} quantifies these properties using Krippendorff's $\alpha$ agreement and Spearman's $\rho$ correlation across the two inference settings (Section~\ref{sec:llm-judgments}). For combined inference---simultaneously judging all utility dimensions---we first analyze bidirectional LLM-to-LLM agreement (first column) to detect prompt order effects. While substantial agreement exists, prompt-induced inconsistencies are present. Further, LLM-to-human agreement (second column) shows only little overlap with gold labels, calling into question the correctness of LLM judgments. This could be due to their high cross-dimensional correlation (third column), indicating detection of a general preference in each comparison, rather than a fine-grained view of individual dimensions. Individual inference, where dimensions are evaluated separately, exhibits similar patterns: while improved LLM-to-LLM agreement suggests improved consistency, LLM-to-human agreement remains poor. Increased dimensional correlation in the individual setting indicates that combined prompting, despite lower consistency, better facilitate dimensional differentiation, possibly due to the model being explicitly made aware of other dimensions to be judged. 

\subsection{Summary}
Crowdsourcing can be a reliable source of judgment data for RAG evaluation when controlling for worker competence. Analysis of order effects and utility dimension interdependence confirmed judgments robustness. Using these validated judgments, we find that LLM-generated responses exhibit significantly higher quality than human-written ones across most utility dimensions. Bullet-style responses are preferred to essay and news styles. However, contrary to judgment-based evaluation, reference-based evaluation is not easily operationalizable, as all tested evaluation metrics demonstrated severe limitations. This suggests that judgment-based approaches are a more promising approach to RAG evaluation. Yet, using LLMs as utility judges, in an effort to improve the efficiency of judgment-based evaluation, fails to produce consistent and correct judgments across all tested settings. 

\section{Conclusion}

We summarize our main findings below, grouped into those relevant to RAG model development, and those relevant to RAG evaluation. We end our paper with a reflection on its limitations and include ethical considerations on study design and attained results.

\paragraph{Findings for RAG Model Development} 

We find response style exerts significant influence on perceived utility of responses; while most current RAG models focus on continuous text, akin to the essay style, bullet-style responses exhibit higher preference judgments. This prompts future work on style adaption for RAG, as specific style instructions can yield stronger preferences from the same model. Furthermore, readability of LLM-generated responses could be improved, scoring worse than human-written responses.

\paragraph{Findings for RAG Evaluation}

Reference-based evaluation fails to accurately operationalize any of the evaluated utility dimensions. This is in line with previous findings~\cite{zhang:2024,goyal:2022}. We thus discourage its use and instead demonstrate that judgment-based evaluation is a feasible and accurate alternative when using crowdsourced data. The increasingly adopted practice of using LLMs as judges of utility, however, is concerning, as they fail to produce consistent and correct judgments in a zero-shot setting. For future work, we will study label transfer from a pool of existing human-judged responses to a new, unjudged responses, akin to a few-shot setting. The utility dimensions previously motivated theoretically~\cite{gienapp:2024} have been shown empirically to be distinguishable by human judges.

\paragraph{Limitations \& Ethical Considerations}

We acknowledge methodological constraints associated with our study. The exclusive use of TREC RAG 2024 topics may limit generalizability. While the utility dimensions we employ aim for comprehensiveness, they may not capture all aspects of RAG system effectiveness relevant to practitioners. The crowdsourcing setup, specifically the worker population, may introduce quality variations and cultural biases in both tasks. Furthermore, our investigation of LLM-written responses relies on a single model configuration, leaving an investigation of other LLM configurations and prompting strategies for future work, which can be compared or validated against our collected data. 

The collection of human-written responses and human judgments raises privacy concerns, which we mitigated through informed consent and comprehensive data anonymization. Crowdsourcing raises concerns about fair labor practices, which we addressed by implementing above-market compensation rates, flexible time allocations, and transparent task descriptions. Overall, our work contributes to a broader discussion on the socio-technical implications of AI-generated responses for search, and highlights the need for RAG evaluation practices grounded in human judgment.

\begin{acks}
This publication has been partially supported by the ScaDS.AI Center for Scalable Data Analytics and Artificial Intelligence, funded by the Federal Ministry of Education and Research of Germany and by the S{\"a}chsische Staatsministerium f{\"u}r Wissenschaft, Kultur und Tourismus; by a Research Fellowship for Harrisen Scells from the Alexander von Humboldt Foundation; and by the OpenWebSearch.eu project, funded by the European Union (GA~101070014). 
\end{acks}

\bibliographystyle{ACM-Reference-Format}
\balance
\bibliography{sigir25-rag-crowdsourcing-lit}

\end{document}